# Comment

**José M. Bernardo**

The authors provide an authoritative lecture guide of *Theory of Probability*, where they clearly state that the more useful material today is that contained in Chapters 3 and 5, which respectively deal with *estimation*, and *hypothesis testing*. We argue that, from a contemporary viewpoint, the impact of Jeffreys proposals on those two problems is rather different, and we describe what we perceive to be the state of the question nowadays, suggesting that Jeffreys's dramatically different treatment is not necessary, and that a joint objective approach to those two problems is indeed possible.

## 1. INTRODUCTION

As the authors point out, *Theory of Probability* is an indispensable, if often difficult to navigate, Bayesian foundational text. Their authoritative lecture guide is therefore very welcome. As should be clear from their review, the main useful material today is contained in Chapters 3 and 5 which, respectively, deal with *estimation*, in the sense of deriving an objective posterior distribution for the quantity of interest, and *hypothesis testing*, presented as a derivation of an objective posterior probability for the hypothesis under consideration. I believe that, from a contemporary viewpoint, the impact of Jeffreys proposals on those two problems is rather different, as I now briefly try to describe.

## 2. ESTIMATION

*One-parameter Jeffreys estimation prior (Jeffreys rule).* Following his own pioneering work (Jeffreys, 1946), the book introduces in Section 3.10 what it is now considered the main meaning of the confusing denomination "Jeffreys prior." Thus, to obtain an objective posterior density for the parameter $\alpha$ of a probability model $f(x|\alpha)$, he proposes the formal use in Bayes theorem of the (often improper) prior $\pi(\alpha) \propto |I(\alpha)|^{1/2}$, where $I(\alpha)$ is Fisher information function. As the authors point out, Jeffreys's motivation is rather obscure: he describes $I(\alpha)$ as a second order approximation to two functional distances, and notes that $|I(\alpha)|^{1/2}$ happens to be invariant under one-to-one transformations. No trace of its more intuitive interpretation in terms of the prior which assigns equal probabilities to equally distinguishable subregions of the parameter space (Lindley, 1961). Also, even in its third (1961) edition, the book only gives a cursory reference to the independent, essentially simultaneous, derivation of the same "rule" produced by Perks (1947) in a much underrated paper. That said, Jeffreys (or Jeffreys–Perks) rule is today the objective prior of choice for regular problems with one continuous parameter, and has been justified in this simple case from many different viewpoints, including coverage properties (Welch and Peers, 1963), minimum bias (Hartigan, 1965), data translation (Box and Tiao, 1973) and information-theoretic arguments (Bernardo, 1979; Berger, Bernardo and Sun, 2009). In one-parameter problems, Jeffreys left without solution non-regular models (e.g., those where the sampling space depends on the parameter) and models with a discrete parameter (although he suggested a very interesting hierarchical argument to deal with the particular example of the hypergeometric distribution).

*Many-parameter Jeffreys estimation prior (multiparameter Jeffreys rule).* The arguments used to propose his rule for one continuous parameter regular models extend to the corresponding multiparameter case, leading to $\pi(\boldsymbol{\alpha}) \propto |I(\boldsymbol{\alpha})|^{1/2}$, where $I(\boldsymbol{\alpha})$ is now Fisher information matrix. As the authors point out in their review, Jeffreys immediately realized, however, that his multivariate rule does not generally produce sensible answers and suggested ad hoc alternatives in virtually all the multiparameter examples he analyzed, leading to a plethora of "Jeffreys priors" in the sense that they were proposed by him, although they do not follow from his general rule. Moreover, as all Bayesians in his time, Jeffreys was working under the assumption that a *unique* objective prior would be appropriate for all inference problems within a multiparameter model. Stein (1959) paradox already suggested that this could not possibly be true, but it was the discovery of the marginalization paradoxes (Dawid, Stone and Zidek, 1973) what definitely established this as a fact, while the reference priors (Bernardo, 1979) provided the first solution to the problem thus created.

*José M. Bernardo is Professor of Statistics at the Facultad de Matemáticas, Universidad de Valencia, 46100-Burjassot, Valencia, Spain e-mail: jose.m.bernardo@uv.es.*

*Proper posteriors.* Scholars have been often surprised at the fact that, when applicable, Jeffreys rule priors (ever in their multiparameter version) typically produce proper posteriors for all data sets, a condition one should certainly require for any proposal of an objective prior to be permissible. I wonder if the authors have an explanation for this remarkable fact (shared by reference priors). Curiously, as noted by the authors, Jeffreys uses in his analysis of the Poisson model the prior $\pi(\alpha) \propto 1/\alpha$ on an scale invariant argument, only to mention later that it is the Jeffreys-rule prior, $\pi(\alpha) \propto 1/\sqrt{\alpha}$, the one leading to a proper posterior for all possible data sets.

## 3. HYPOTHESIS TESTING

*Jeffreys hypothesis testing priors.* In Chapter 5, Jeffreys focuses on precise hypothesis testing and, as the present review indicates, does not produce a solution with the level of acceptance and generality of his one-parameter rule for estimation. Jeffreys intends to obtain a posterior probability for a precise null hypothesis and, to do this, he is forced to use a mixed prior which puts a lump of probability $p = \Pr(H_0)$ on the null, say $H_0 \equiv \{\theta = \theta_0\}$, and distributes the rest with a *proper* prior $p(\theta)$ (he mostly chooses $p = 1/2$). This has a very upsetting consequence, usually known as Lindley's paradox (Lindley, 1957): for any fixed prior probability $p$ independent of the sample sixe $n$, the procedure will wrongly accept $H_0$ whenever the likelihood is concentrated around a true parameter value which lies $O(n^{-1/2})$ from $H_0$. I find it difficult to accept a procedure which is *known* to produce the wrong answer under specific, but not controllable, circumstances; see Robert (1993) for a relatively recent review of this fascinating issue. Besides this, I believe, serious problem, Jeffreys suggestion of a Cauchy density for the required *proper* prior is rather *ad hoc* and does not generalize to more complicated problems. There have been many attempts to define priors intended to obtain objective posterior probabilities for precise nulls. To the best of my knowledge, none of those has emerged as a clearly acceptable general solution.

*Hypothesis testing with conventional reference priors.* To test whether or not a precise hypothesis $H_0$ is compatible with observed data, it is not necessary to try to obtain a posterior probability for $H_0$, and hence it it not necessary to use a totally different type of objective prior than that used for estimation. As forcefully argued by Jaynes (1980), all that is required is to obtain the posterior distribution of a quantity which measures the discrepancy between the true model and the null model. A very attractive candidate is the *intrinsic discrepancy*. The intrinsic discrepancy between two probability distributions $p_1$ and $p_2$ for **x** is defined as

$$\delta\{p_1, p_2\} = \min[\kappa\{p_1|p_2\}, \kappa\{p_2|p_1\}],$$

where $\kappa\{p_j|p_i\} = \int_{\mathcal{X}_i} p_i(\mathbf{x}) \log[p_i(\mathbf{x})/p_j(\mathbf{x})] \, d\mathbf{x}$, the Kullback–Leibler (KL) divergence of $p_j$ from $p_i$. This inherits all the very nice properties of the KL divergence (non-negative, invariant, additive) but is also symmetric and it is defined even if the supports of the two distributions are strictly nested. For instance, in the canonical example of testing whether a random sample $\mathbf{x} = \{x_1, \ldots, x_n\}$ from a normal $N(x|\mu, \sigma)$ is or is not compatible with the mean value $\mu_0$, one obtains

$$\delta\{\mu_0, (\mu, \sigma)\} = \delta\{N(x|\mu_0, \sigma), N(x|\mu, \sigma)\}$$
$$= \frac{1}{2}\left(\frac{\mu - \mu_0}{\sigma/\sqrt{n}}\right)^2.$$

If $\sigma$ is known, the objective posterior distribution of $\delta\{\mu_0, (\mu, \sigma)\}$ with the usual objective prior $\pi(\mu) = 1$ gives all required information about whether or not the null $H_0 \equiv \{\mu = \mu_0\}$ should be accepted, including the size of the plausible departures. If a formal decision is required, $\delta\{\mu_0, (\mu, \sigma)\}$ may be used as a loss function (it is an intrinsic loss in the sense of Robert, 1996). In this case, one simply computes its expected value, which is the intrinsic test statistic

$$d(\mu_0|\mathbf{x}) = \frac{1}{2}(1 + z^2), \quad z = \frac{\bar{x} - \mu_0}{\sigma/\sqrt{n}},$$

and rejects the null if this is too large (say larger than log[100] since this would imply that the data are expected to be over 100 times more likely under the true model than under the null model). See

Bernardo and Rueda (2002) and Bernardo (2005) for the general definition when there are nuisance parameters, and for many specific examples. One could certainly use other continuous loss functions but the point is, Bayesian testing of precise nulls do *not* necessarily require the use of mixed priors as those suggested by Jeffreys for this problems, and this has the nontrivial merit of being able to use for both estimation and hypothesis testing problems a single, unified theory for the derivation of objective "reference" priors.

## ACKNOWLEDGMENT

Work supported in part by MEC Grant MTM2006-07801, Spain.